\documentclass[9pt, twocolumn,twoside]{IEEEtran}
\usepackage{amsmath,graphicx}

\usepackage{amsmath,amsfonts} 
\usepackage{amssymb, amsthm}
\usepackage{graphicx}
\usepackage{booktabs}
\usepackage{multirow}
\usepackage{cite}
\usepackage{color}
\usepackage{microtype}
\usepackage{nicefrac}

\renewcommand{\eqref}[1]{(\ref{#1})}
\newcommand{\secref}[1]{\mbox{Section~\ref{#1}}}
\newcommand{\figref}[1]{\mbox{Figure~\ref{#1}}}

\newcommand{\setO}{\ensuremath{\mathcal{O}}}

\newcommand{\bmd}{\ensuremath{\mathbf{d}}}

\newcommand{\bmn}{\ensuremath{\mathbf{n}}}

\newcommand{\bmdbar}{\ensuremath{\bar{\bmd}}}

\newcommand{\bA}{\ensuremath{\mathbf{A}}}

\newcommand{\bF}{\ensuremath{\mathbf{F}}}
\newcommand{\bG}{\ensuremath{\mathbf{G}}}
\newcommand{\bH}{\ensuremath{\mathbf{H}}}
\newcommand{\bI}{\ensuremath{\mathbf{I}}}

\newcommand{\bDbar}{\ensuremath{\bar{\mathbf{D}}}}

\newcommand{\MT}{{\ensuremath{U}}}
\newcommand{\MR}{{\ensuremath{B}}}



\IEEEoverridecommandlockouts

\title{Linear Large-Scale MIMO Data Detection \\ for 5G Multi-Carrier Waveform Candidates}

\author{
    \IEEEauthorblockN{Nihat Engin Tunali$^\text{1}$, Michael Wu$^\text{1,2}$, Chris Dick$^\text{3}$, and Christoph Studer$^\text{2}$} \\[0.3cm]   
    \IEEEauthorblockA{$^\text{1}$Xilinx Inc, San Jose, CA; e-mail: \{engint,\,chrisd\}@xilinx.com} \\
    \IEEEauthorblockA{$^\text{2}$Rice University, Houston, TX; e-mail: mbw2@rice.edu} \\
    \IEEEauthorblockA{$^\text{3}$School~of ECE, Cornell University, Ithaca, NY; e-mail: studer@cornell.edu} \\
\thanks{The work of C. Studer was supported by Xilinx Inc.\ and by the US National Science Foundation (NSF) under grants ECCS-1408006 and CCF-1535897.}
}

\begin{document}

\maketitle

\sloppy

\begin{abstract}
Fifth generation (5G) wireless systems are expected to combine emerging transmission technologies, such as large-scale multiple-input multiple-output (MIMO) and non-orthogonal multi-carrier waveforms, to improve the spectral efficiency and to reduce out-of-band (OOB) emissions. This paper investigates the efficacy of two promising multi-carrier waveforms that reduce OOB emissions in combination with large-scale MIMO, namely filter bank multi-carrier (FBMC) and generalized frequency division multiplexing (GFDM). We develop  novel, low-complexity data detection algorithms for both of these waveforms. We investigate the associated performance/complexity trade-offs in the context of large-scale MIMO, and we study the peak-to-average power ratio (PAPR). Our results show that reducing the OOB emissions with FBMC and GFDM leads to higher computational complexity and PAPR compared to that of orthogonal frequency-division multiplexing (OFDM) and single-carrier frequency division multiple access (SC-FDMA).
\end{abstract}
\section{Introduction}

Fifth generation (5G) wireless technologies are driven by higher data rates (in the order of 10\,Gb/s) and lower latency (in the order of 1\,ms) compared to that of existing standards. In order to address these requirements, large-scale multiple-input multiple-output (MIMO) has been proposed and is currently an active area of research \cite{Rusek2012,larsson2014massive}. In addition to higher data rates, 5G is also expected to support a variety of applications that range from bursty machine-to-machine (M2M) type communications, to high bandwidth video streaming. Supporting all these applications with closely packed or potentially overlapping bandwidths necessitates efficient utilization of the frequency spectrum; this requires the deployment of novel waveforms that exhibit strong frequency localization and hence, low out-of-band (OOB) emissions.  

In the past years, two potential waveform candidates have emerged in order to address the strong frequency-localization requirements of 5G. One of these waveforms is generalized frequency division multiplexing (GFDM) \cite{NicolaFettweiss2012}, a generalization of OFDM in which individual sub-carrier signals are circularly convolved with a prototype filter that improves frequency localization. The second waveform candidate is filter-bank multi-carrier (FBMC) \cite{siohan2002analysis}, in which offset quadrature amplitude modulation is employed and individual sub-carrier signals are passed through a poly-phase filter network, which imposes strong frequency localization and can be implemented efficiently with filter banks.
Despite the fact that GFDM and FBMC address the reduced OOB-emission requirements of 5G, the implications  of using these waveforms on error-rate performance, computational complexity, and linearity requirements in large-scale MIMO systems with potentially hundreds of antennas at the base station are unclear.

\subsection{Contributions}
In this paper, we propose novel, low complexity data detection algorithms for GFDM and FBMC, and we analyze their efficacy in the context of large-scale MIMO. 
We show frame error-rate  (FER) simulation results and evaluate the associated computational complexity in terms of operation counts of GFDM and FBMC for various antenna configurations. We furthermore study the associated linearity requirements in terms of the peak-to-average power ratio (PAPR) and investigate the associated OOB emissions. We finally compare these waveform candidates to orthogonal frequency division multiplexing (OFDM) and single-carrier frequncy division multiple access (SC-FDMA), which are used in the downlink and the uplink of 3GPP LTE, respectively~\cite{SesiaLTE}.

\subsection{Notation}
Lowercase boldface letters denote column vectors and uppercase boldface letters denote matrices. For a matrix  $\mathbf{A}$, we denote its $j$th column as $\mathbf{a}_j$, and its entry at the $i$th row and $j$th column as $a_{i,j}$; we denote the transpose and the Hermitian transpose as $\mathbf{A}^T$ and $\mathbf{A}^H$, respectively. We denote the $N\times N$ identity matrix, normalized forward and inverse discrete Fourier transform (DFT) matrix as $\mathbf{I}_N$, $\mathbf{F}_N$, and $\mathbf{F}_N^H$, respectively, and we have  $\mathbf{F}_N\mathbf{F}_N^H=\mathbf{I}_N$. For a column vector $\mathbf{a}$, we denote the $i$th entry as $a_i$. For a given column vector $\mathbf{v}$, we denote $1/\mathbf{v}$ as the vector that consists of element-wise reciprocals of $\mathbf{v}$.  We denote the complex conjugate of a scalar $s$ by $s^{*}$.

\section{System Model}
\label{sec:system}
\subsection{Uplink System Model}

We consider the uplink of a large-scale multi-user MIMO (MU-MIMO) system where $\MT$ single-antenna users transmit data to~$\MR$ BS antennas (typically larger than $\MT$).  
The $i$th user transmits $M$ frequency domain~(FD) blocks of symbols $\mathbf{s}^{(i)} = \big[\mathbf{s}^{(i)}_0,\ldots, \mathbf{s}^{(i)}_{M-1}\big]$, where the $K$-dimensional entries of the $m$th block $\mathbf{s}^{(i)}_m  = \big[s^{(i)}_{m,0},\ldots,s^{(i)}_{m,K-1}\big]^T$ are assigned to $K$ dedicated data-carrying subcarriers. 

We model the received FD symbols on the $k$th  subcarrier for the $m$th block  as $\mathbf{y}_{k,m} = \bH_{k,m}\mathbf{s}_{k,m} +\bmn_{k,m}$. Here,  the receive vector, channel matrix, transmit vector, and noise vector defined as follows:
\begin{align*}
\mathbf{y}_{k,m}&=[y_{k,m}^{(0)},\ldots,y_{k,m}^{(B-1)}]^T.
\end{align*}
\begin{align*}
\bH_{k,m} = \left(\begin{array}{ccc}
H^{(0,0)}_{k,m} & \cdots & H^{(0,\MT -1)}_{k,m}\\[-0.2cm]
\vdots & \ddots & \vdots\\
H^{(\MR -1,0)}_{k,m} & \cdots & H^{(\MR -1,\MT -1)}_{k,m}
\end{array}\right)\!,
\end{align*}
\begin{align*}
\mathbf{s}_{k,m}&= [s_{k,m}^{(0)},\ldots,s_{k,m}^{(U-1)}]^T,\\ 
\bmn_{k,m}&=[n_{k,m}^{(0)},\ldots,n_{k,m}^{(B-1)}]^T.
\end{align*}
For the $m$th block, $y^{(i)}_{k,m}$ denotes the FD symbol received on the $k$th subcarrier for antenna $i$, $H^{(i,j)}_{k,m}$ is the corresponding (flat-fading) frequency gain on the $k$th subcarrier between the $i$th receive antenna and $j$th user. The scalar $s^{(j)}_{k,m}$ denotes the symbol transmitted by the $j$th user on the $k$th subcarrier of the $m$th block; the scalar $n^{(i)}_{k,m}$ represents i.i.d.\ complex  circularly symmetric Gaussian noise with variance $N_0$. 

The multi-carrier modulation schemes considered in the remainder of the paper use different methods for constructing the $K$-length FD blocks. For OFDM modulation, we have $\mathbf{s}^{(i)}_m  = \bmd^{(i)}_m$, where $\bmd^{(i)}_m =  \big[d^{(i)}_{m,0},\ldots,d^{(i)}_{m,K-1}\big]^T$. An element of  $\bmd^{(i)}_m$, $d^{(i)}_{m,k}$, is a constellation point drawn from a finite alphabet~$\setO$~(such as QPSK) with unit average transmit power. To reduce the typically high peak-to-average power ratio (PAPR) of OFDM, SC-FDMA modulation precodes the transmit vector using a DFT, where~$\mathbf{s}^{(i)}_m  = \bF_K{\bmd^{(i)}_m}$. To improve the  frequency localization compared to that of SC-FDMA and OFDM, both GFDM and FBMC require additional steps to construct $\mathbf{s}^{(i)}_m$. We detail the GFDM modulator in~\secref{GFDM:modulator} and the FBMC modulator in~\secref{FBMC:modulator}. Since modulation is carried out on a per-user basis, we omit the user index $i$ in the following to simplify notation. Furthermore, GFDM and FBMC modulate the data symbols across both the subcarriers and subsymbols, and we describe the modulation process in time domain (TD). In particular, we denote the GFDM and FBMC modulated signals as  $\mathbf{x}=[(\mathbf{F}^H \mathbf{s}_0)^T,(\mathbf{F}^H \mathbf{s}_1)^T,\ldots, (\mathbf{F}^H \mathbf{s}_{M-1})^T]^T$.

\subsection{GFDM Modulator}
\label{GFDM:modulator}
\begin{figure}[tp]
\centering
\includegraphics[width=0.90\columnwidth]{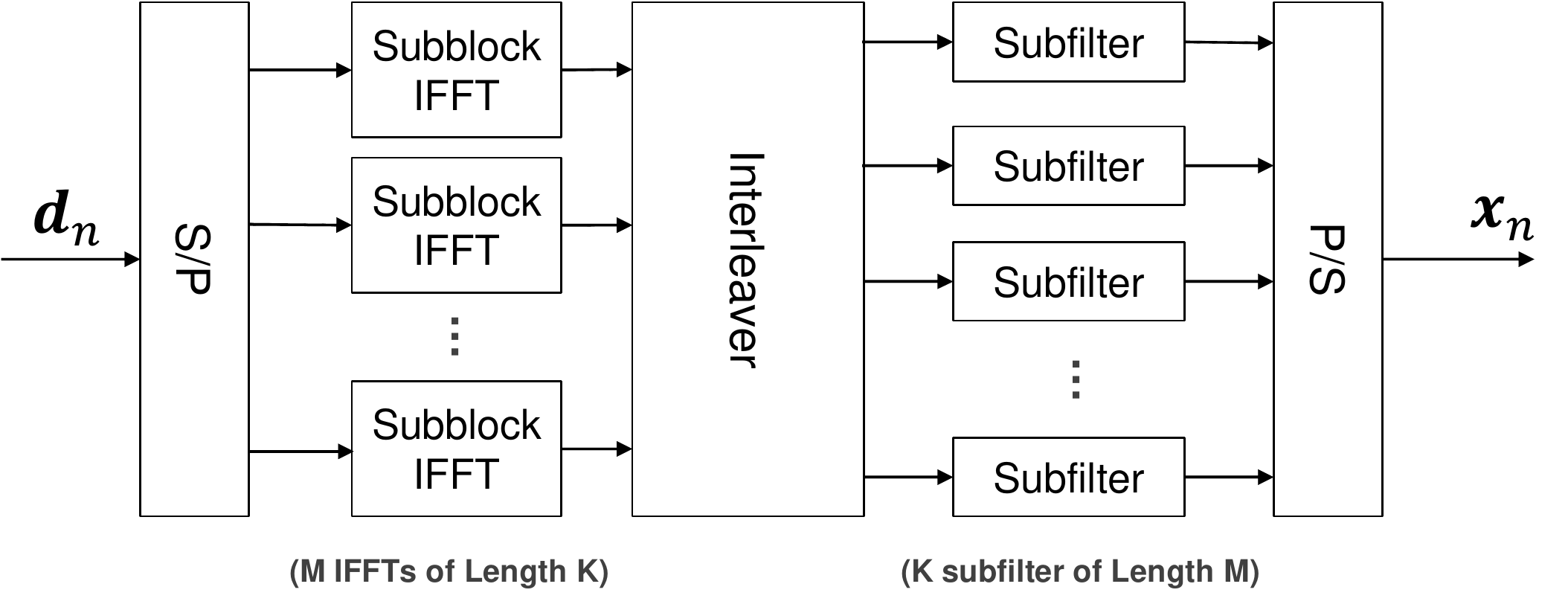}
\caption{Overview of the considered low complexity GFDM modulator.}\label{Fig:2}
\vspace{-0.3cm}
\end{figure}

A GFDM modulator performs per-subcarrier filtering in order to reduce OOB interference (or emissions). 
We consider the low complexity GFDM modulator proposed in~\cite{farhang2015low}.  To this end, we first rearrange the QAM data symbols $d_{m,k}$ into a matrix $\mathbf{D}=[\mathbf{d}_0,\ldots,\mathbf{d}_{M-1}]$, where $\mathbf{d}_{m}=[d_{m,0},\ldots, d_{m,K-1}]^T$. We then take the $K$-point DFT of each column of $\mathbf{D}$ and transpose the resulting matrix, i.e.,  $\bDbar=[\mathbf{F}_K\mathbf{d}_0,\ldots,\mathbf{F}_K\mathbf{d}_{M-1}]^T$. 
Suppose that $\mathbf{g}$ is a prototype filter (typically chosen as a root-raised cosine filter) of length~$MK$.
We then circularly convolve each of the columns $\bmdbar_k$ of $\bar{\mathbf{D}}$ with the corresponding polyphase components of the transmit filter and obtain the GFDM signal in the TD as follows:
\begin{align}\label{eq:2}
\mathbf{x}_k=\mathbf{g}_k\circledast \bar{\mathbf{d}}_k,
\end{align}
where we used 
\begin{align*}
\mathbf{x}_k&=\left[x_k,x_{k+K},\ldots, x_{k+(M-1)K}\right]^T\\
\mathbf{g}_k&=\left [g_k,g_{k+K},\ldots ,g_{k+(M-1)K}\right]^T.
\end{align*}
The low-complexity GFDM modulator is depicted in~\figref{Fig:2}.

\subsection{FBMC Modulator}
\label{FBMC:modulator}
Suppose that our data symbols $d_{m,k}$ are chosen from a pulse amplitude modulation (PAM) constellation. FBMC modulation on this set of data symbols can be expressed compactly by the following equation~\cite{siohan2002analysis}:
\begin{align}\label{eq:4}
x_n=\sum_{m=0}^{M-1}p_{n-m\frac{K}{2}}\sum_{k=0}^{K-1}d_{m,k}e^{ \frac{j2\pi kn}{K}}e^{\frac{-j2\pi k(\frac{L-1}{2})}{K}}j^{m+k}.
\end{align}
Here, $p_n$ denotes a prototype filter and $L$ denotes its length. The choice of the prototype filter is crucial in order to mitigate inter-symbol interference (ISI) and inter-carrier interference (ICI) for FBMC, and there have been numerous related studies~\cite{FBMCfilterdes}. We use the well-known PHYDYAS filter~\cite{bellanger2010fbmc} of length $L=4K$ since this filter almost completely eliminates ISI and ICI for FBMC and imposes strong frequency localization. Unlike GFDM, the filtering operation is carried out through a standard convolution in FBMC and therefore, the modulated signal $x_n$ is of length  $K(M+L-1/2)$.

\begin{figure}[tp]
\centering
\includegraphics[width=0.95\columnwidth]{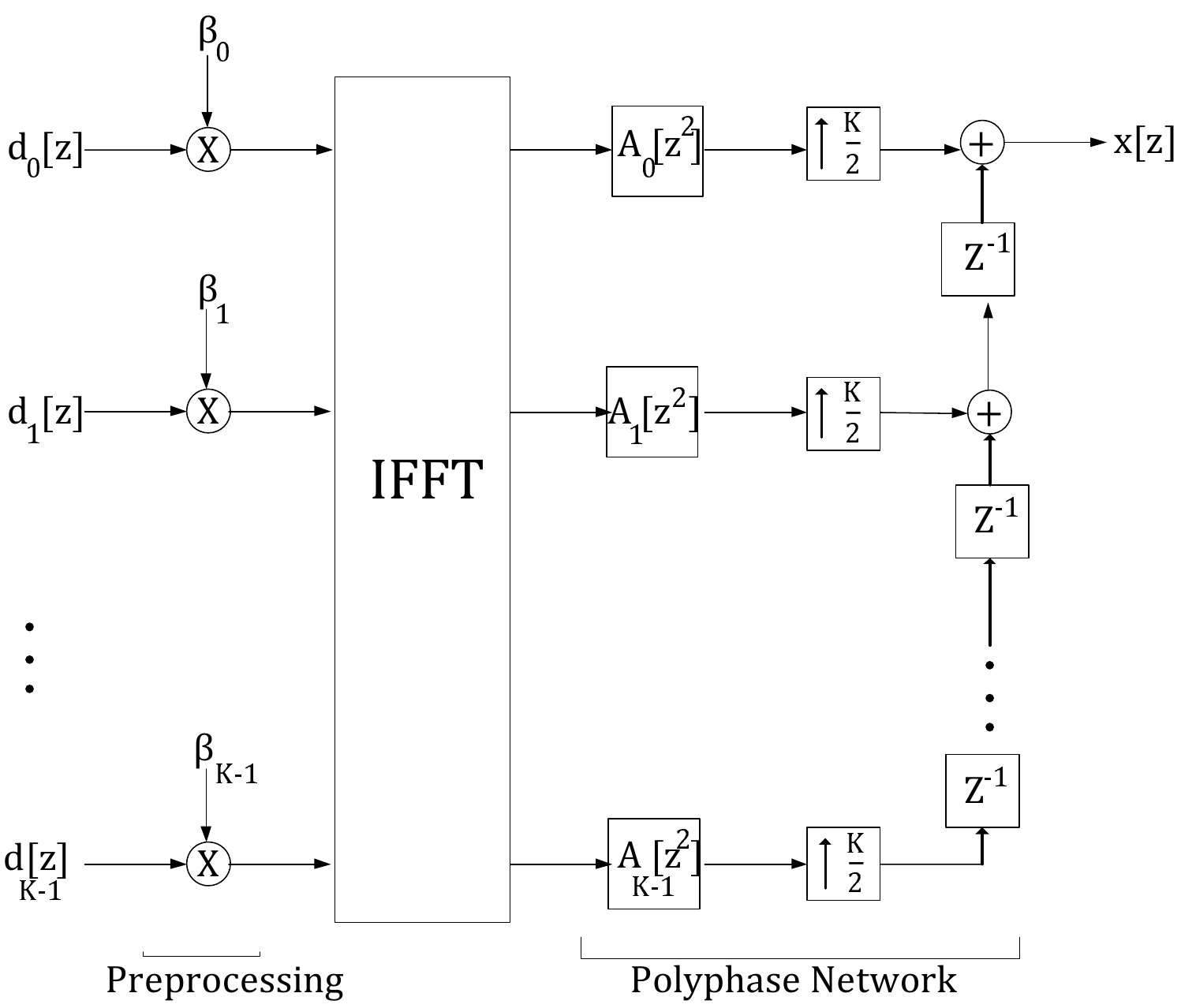}
\caption{Overview of the considered low-complexity FBMC modulator.}
\label{Fig:4}
\vspace{-0.3cm}
\end{figure}

 By inspecting the right-hand side of~\eqref{eq:4}, we see that the quantities $d_{m,k}$ are scaled by $e^{ \frac{-j2\pi k({L-1})}{2K}}j^{m+k}$ and a $K$-point IDFT is taken across all the sub-carriers of a given subsymbol. Then, as seen in the left-hand side of~\eqref{eq:4}, the subsymbols are filtered with the prototype filter, which can be implemented efficiently with a polyphase network. Define
\begin{align*}
d_k[z] & \textstyle =\sum_{m=0}^{M-1}d_{m,k}z^{-m},\\
A_k[z] & \textstyle =\sum_{l=0}^{L-1}p_{k+lK}z^{-l},\\
\beta_k& \textstyle =e^{ \frac{-j2\pi k({L-1})}{2K}}j^{m+k},
\end{align*}
where $d_k[z]$ are the subsymbols transmitted on sub-carrier $k$ represented in the Z-domain and $A_k[z]$ are the polyphase filter components of the prototype filter represented in the z-domain. A well-known low complexity implementation of FBMC modulation~\cite{siohan2002analysis, nadal2014hardware} in the z-domain is depicted in~\figref{Fig:4}.
Note that the upsampling by $\nicefrac{K}{2}$ in~\figref{Fig:4} takes place due to filtering with $p_{n-\frac{mK}{2}}$ as seen in~\eqref{eq:4}

\section{Linear Data Detection}

\subsection{Linear Frequency Domain Equalizer}

To arrive at low-complexity data detection, we exclusively focus on linear FD-MMSE equalization methods that operate on a per-subcarrier basis for each block. To this end, we compute the regularized Gram matrix $\bA_{k,w}=\bG_{k,w}+N_0\bI_\MT$ and the matched filter (MF) vector $\mathbf{f}^\text{MF}_{k,w} = \bH_{k,w}^H \mathbf{y}_{k,w}$. To compute the FD-equalized symbols, we compute $\hat{\mathbf{s}}_{k,w} = \bA_{k,w}^{-1}\mathbf{y}^\text{MF}_{k,w}$.

Given the FD-equalized symbols for the $i$th user, $\hat{\mathbf{s}}^{i}$, the demodulator goes back to the time domain and computes the soft-output information in the form of log-likelihood ratio (LLR) values for the transmitted bits from $\mathbf{\hat{x}}=[(\mathbf{F}^H \mathbf{\hat{s}}_0)^T,(\mathbf{F}^H \mathbf{\hat{s}}_1)^T,\ldots, (\mathbf{F}^H \mathbf{\hat{s}}_{M-1})^T]^T$. Note that once we go back to the TD, the noise plus interference (NPI) variance needs to be computed. In \cite{Wu2014}, the NPI variance was derived for SC-FDMA. Since each modulation scheme processes the transmitted data bits differently, the corresponding demodulators perform different steps. While there are existing, low-complexity algorithms for OFDM and SC-FDMA~\cite{Studer2011, Wu2014}, we develop new data detection algorithms for GFDM and FBMC and derive the NPI variance computation for these waveforms. Since modulation is carried out on a per-user basis, we omit the user index $i$ in the following.

\subsection{Linear GFDM Receivers}\label{GFDM: ZF}

In literature, three linear data detectors for GFDM have been proposed: matched filter (MF), zero forcing (ZF), and minimum mean square error (MMSE)  \cite{NicolaFettweiss2012,farhang2015low}. As expected, it was observed that the MF-GFDM receiver performs worst in terms of error-rate performance, whereas the ZF-GFDM and MMSE-GFDM receivers provide comparable performance~\cite{NicolaFettweiss2012}. In this paper, we focus on the ZF-GFDM receiver, as there is only a small penalty in error-rate performance compared to the MMSE-GFDM receiver. Furthermore, a low-complexity algorithm for ZF-GFDM requires fewer operations than the MMSE-GFDM receiver~\cite{farhang2015low}.

\begin{figure}[tp]
\centering
\includegraphics[width=3.5in]{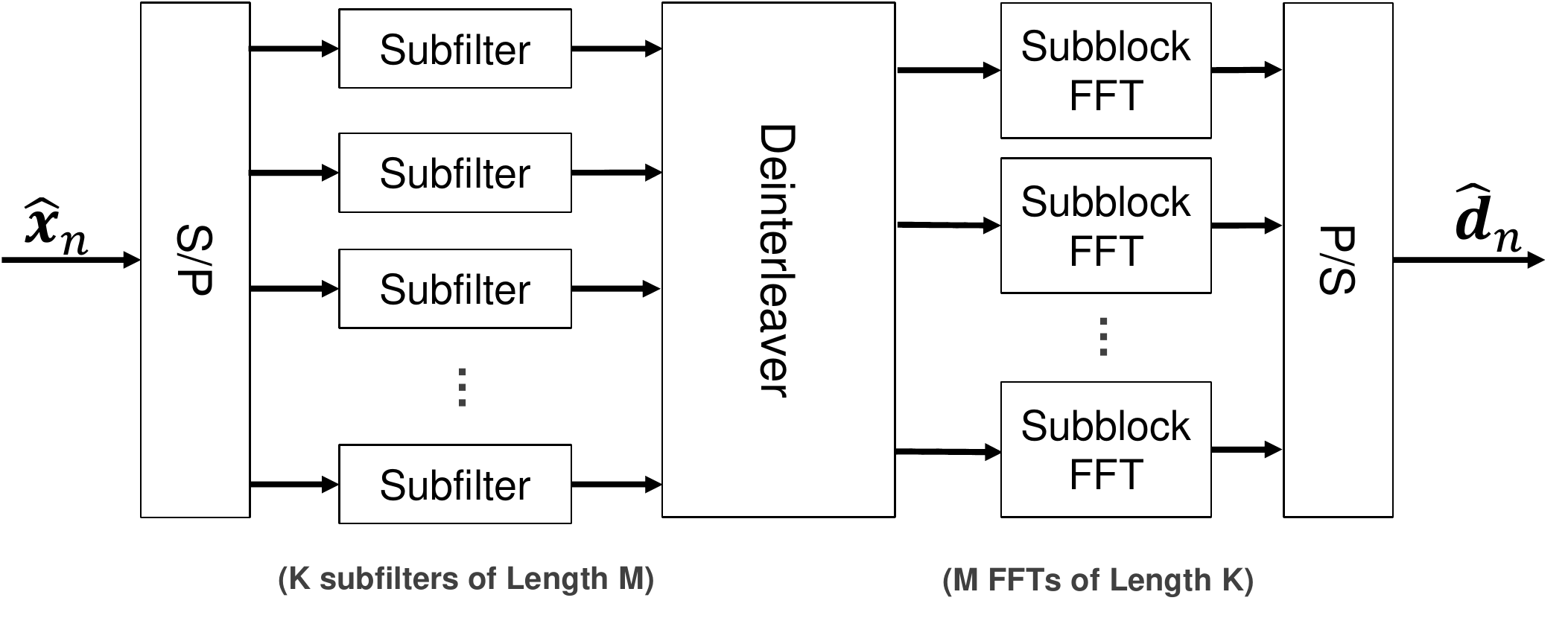}
\caption{Overview of the low-complexity ZF-GFDM data detector.}\label{Fig:3}
\vspace{-0.3cm}
\end{figure}

In order to implement a zero-forcing receiver in GFDM, a straightforward matrix/vector multiplication is computationally demanding as it requires $(KM)^2$ complex-valued multiplications. Suppose that our FD equalized GFDM symbols in time domain are $\mathbf{\hat{x}}$.  A low-complexity ZF-GFDM receiver can be implemented by reversing the modulation steps in \ref{GFDM:modulator} as follows.
Let $\mathbf{\hat{x}}_k=\left[\hat{x}_k,\hat{x}_{k+K},\ldots,\hat{x}_{k+(M-1)K}\right]^T$ and $\tilde{\mathbf{g}}_k$ be chosen such that  $\mathbf{F}_M\tilde{\mathbf{g}}_k=1/\mathbf{F}_M{\mathbf{g}}_k$, where $\mathbf{g}_k$ was specified in \eqref{eq:2}. We first circularly convolve $\mathbf{\hat{x}}_k$ with $\tilde{\mathbf{g}}_k$ and obtain
\begin{equation}
{\mathbf{e}}_k=\tilde{\mathbf{g}}_k\circledast {\mathbf{\hat{x}}}_k.
\end{equation}
We insert each ${\mathbf{e}}_k$ as column vectors into a matrix $\mathbf{E}$ and take the transpose of ${\mathbf{E}}$. Denote $\tilde{\mathbf{D}}={\mathbf{E}}^T$. Then, we take the $K$-point IDFT of each column in $\tilde{\mathbf{D}}$ and obtain the estimates of the data symbols as follows:
\begin{align*}
\hat{\mathbf{D}}=[\mathbf{F}^H_K\tilde{\mathbf{d}}_0, \mathbf{F}^H_K\tilde{\mathbf{d}}_1,\ldots,\mathbf{F}^H_K\tilde{\mathbf{d}}_{M-1}].
\end{align*}
The low-complexity ZF-GFDM receiver is shown in~\figref{Fig:3}.

\subsection{NPI Variance computation for GFDM}

In order to compute LLR values \cite{Wu2014}, we need to compute the NPI variance that results from ZF-GFDM equalization. Suppose that after frequency domain MMSE equalization, the TD NPI variance is computed as $v^2$. Recall that in~\secref{GFDM: ZF}, we first circularly convolve the equalized subsymbols $\mathbf{\hat{x}}_k$ of each subcarrier with $\tilde{\mathbf{g}}_k$. Therefore, the NPI variance for sub-carrier $k$, which we denote by $\tilde{v}^2_{k}$, can be computed as follows: 
\begin{equation}
\textstyle \tilde{v}^2_{k}=\frac{1}{M}{v}^2\sum_{m=0}^{M-1}|\tilde{g}_{k,m}|^2,
\end{equation}
where we ignore correlation in the post FD-equalized noise.

Then, since we perform a $K$-point IDFT across each subcarrier of a subsymbol, the final NPI variance, which we denote as $\hat{v}^2$, can be computed as 
\begin{align*}
\textstyle \hat{v}^2=\frac{1}{K}\sum_{k=0}^{K-1}\tilde{v}^2_{k}={v}^2_{k}\frac{1}{KM}\sum_{k=0}^{K-1}\sum_{m=0}^{M-1}|\tilde{g}_{k,m}|^2
\end{align*}
We emphasize that the term $\frac{1}{KM}\sum_{k=0}^{K-1}\sum_{m=0}^{M-1}|\tilde{g}_{k,m}|^2$ can be computed offline. 
\subsection{Linear FBMC Receiver}\label{Sec: FBMC receiver}
\begin{figure}[tp]
\centering
\includegraphics[width=0.95\columnwidth]{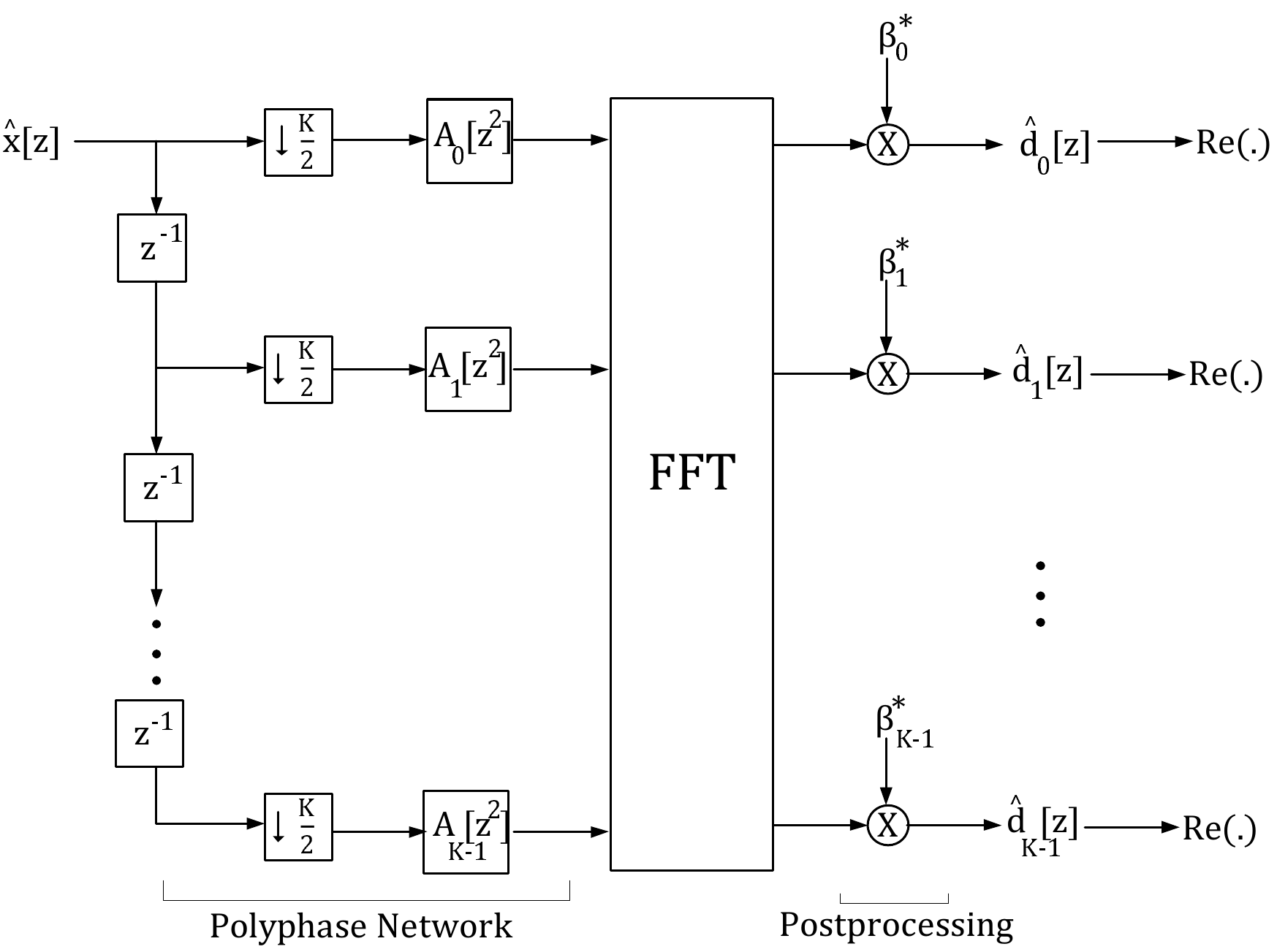}\caption{Overview of the low-complexity FBMC data detector.}\label{Fig:5}
\vspace{-0.3cm}
\end{figure}

Suppose that our FD-equalized FBMC symbols in time domain are denoted by $\mathbf{\hat{x}}$. A low-complexity FBMC data detector can be implemented by reversing the steps in~\secref{FBMC:modulator}, i.e., we pass $\mathbf{\hat{x}}$ through the polyphase network, take a DFT across subcarriers, multiply with $\beta_k^{*}$, and then take the real component. The procedure is illustrated in~\figref{Fig:5} .

\subsection{NPI Variance computation for FBMC}
\begin{figure}[tp]
\centering
\vspace{-3cm}
\includegraphics[width=0.75\columnwidth]{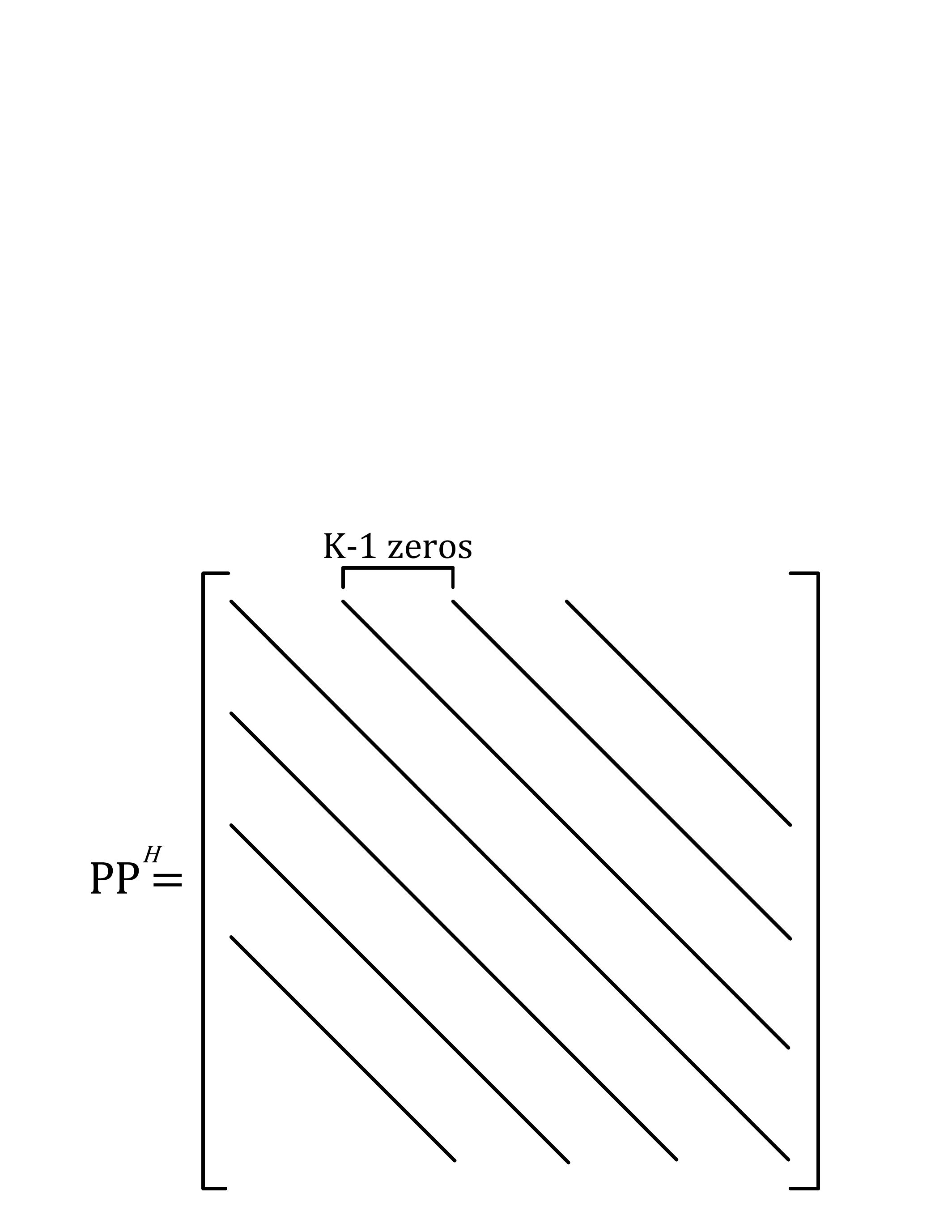}\caption{Structure of the noise covariance matrix of the FMBC receiver.}\label{Fig:6}
\vspace{-0.4cm}
\end{figure}

Suppose that from FD equalization, the NPI variance is computed as $v^2$. Recall that in~\secref{Sec: FBMC receiver}, the first step of the FBMC receiver is to pass the received symbols $\mathbf{\hat{x}}$ through a polyphase network. This operation can be expressed as $\mathbf{P}\mathbf{\hat{x}}$, where $\mathbf{P}$ is a $KM\times K(M+L-1/2)$ sparse matrix that represents the polyphase network. After passing $\mathbf{\hat{x}}$ through a polyphase network, the resulting noise covariance matrix of the FBMC symbols can be expressed as $v^2\mathbf{P}\mathbf{P}^H$. By taking a closer look at the polyphase network, one can observe that $\mathbf{P}\mathbf{P}^H$ is a sparse multi-diagonal matrix as illustrated in~\figref{Fig:6}, where there are $2L-1$ diagonal components and each consecutive diagonal is separated by $K-1$ zeros.

Let $\mathbf{q}$ denote the main diagonal of $\mathbf{P}\mathbf{P}^H$. The next step in the FBMC demodulator is an FFT across the subcarriers of each subsymbol. Due to the gap of $K-1$ zeros between consecutive diagonals in $\mathbf{P}\mathbf{P}^H$, the $K$-point IFFT step averages $K$ entries of $\mathbf{q}$. Hence, the NPI variance vector of subsymbol $m$, which we denote as $\hat{{v}}_{m}^2$, can be computed as 
\begin{align*}
\textstyle \hat{{v}}_{m}^2={v}^2\frac{1}{K}\sum_{k=0}^{K-1} q_{k+mK}.
\end{align*}
We note that the term $\frac{1}{K}\sum_{k=0}^{K-1} q_{k+mK}$ is not data dependent and hence, can be computed offline.
\section{Results}
We next provide simulation results for a $20$\,MHz bandwidth, $K=1200$ subcarrier MU-MIMO LTE uplink scenario, as defined by the LTE specification \cite{3GPPLTE}. For OFDM, SC-FDMA, and GFDM, we assume 64-QAM and ${M=14}$ blocks per frame. For GFDM we use a root-raised cosine filter of length $MK$ with a roll-off factor of $0.25$. For FBMC, we use the PHYDSAS filter \cite{bellanger2010fbmc} of length $4K$; we also assume the data symbols to be drawn from 8-PAM. In order to match the data rate, we assumed ${M=17.5}$ blocks per frame for FBMC. To evaluate the error-rate performance, we use the WINNER-Phase-2 channel model~\cite{winner2}, assume a linear antenna array with antenna spacing of $6$\,cm, and use a per-user rate-$3/4$ 3GPP-LTE turbo code.

\subsection{Complexity  Comparison}
\begin{figure}[tp]
\centering
\includegraphics[width=0.8\columnwidth]{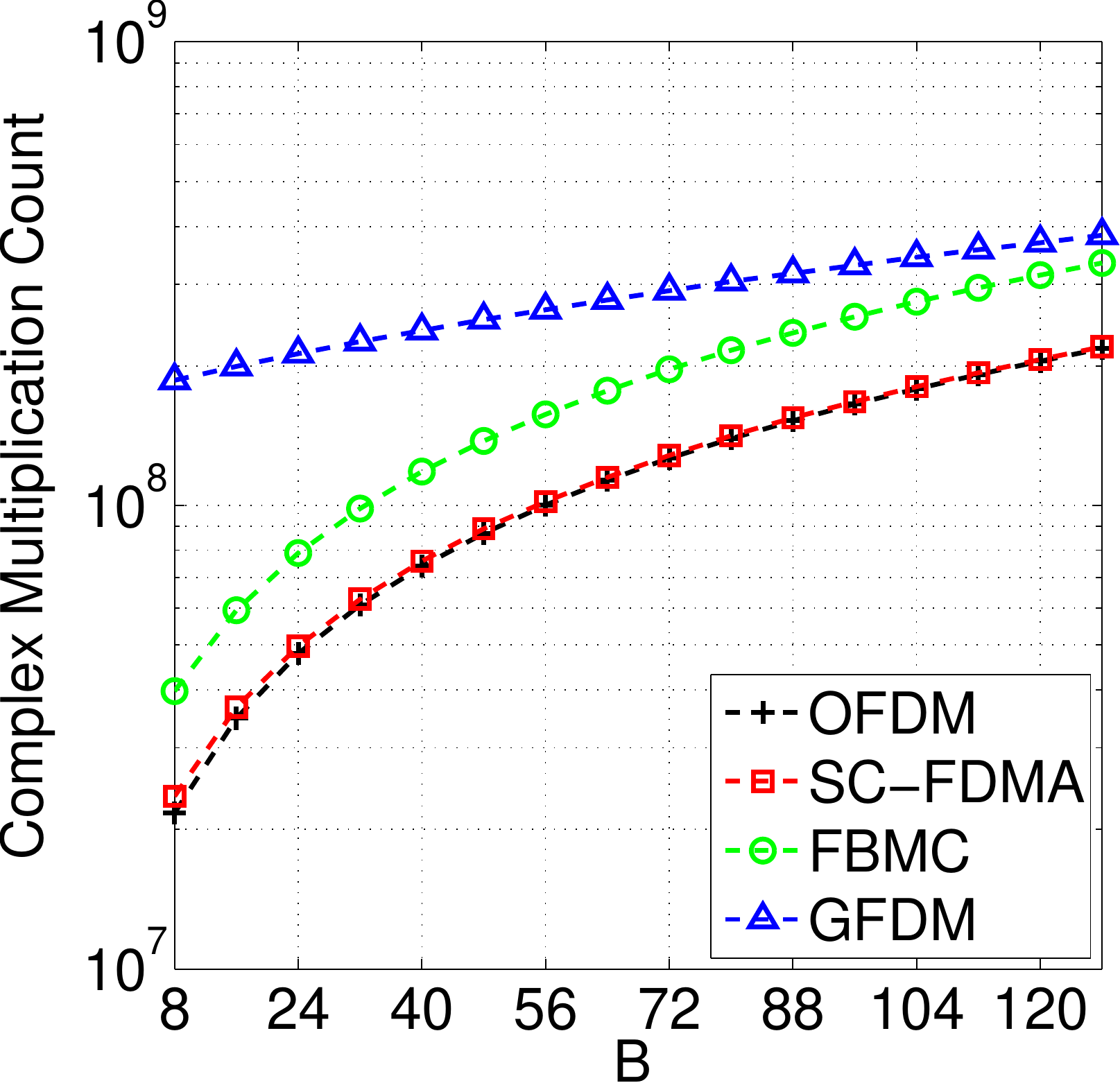}
\caption{Number of complex-valued multiplications for $U=8$ user terminals. OFDM and SC-FDMA exhibit significantly lower complexity than FBMC and GFDM, especially for  small BS-to-user antenna ratios.}\label{Fig:7} 
\vspace{-0.3cm}
\end{figure}

We present a complexity comparison by counting the number of complex-valued multiplications between GFDM, FBMC, OFDM, and SC-FDMA for a $B\times 8$ large-scale MIMO system in~\figref{Fig:7}. As expected, GFDM and FBMC require higher complexity than OFDM and SCFDMA due to the additional filtering  operations. Note that for GFDM, the difference is more prominent since we use a  root-raised cosine filter of length $14K$, whereas for FBMC the PHYDSAS filter is only of length $4K$. We note that for small $B/U$ ratios, the complexity difference is in agreement with other comparisons, such as the ones in \cite{farhang2015low,nadal2014hardware} for a single input single output setup. For larger BS-to-user antenna ratios, the complexity difference between GFDM and other waveforms decreases since the complexity is dominated by FD equalization.

\subsection{PAPR Comparison}
\begin{figure}[tp]
\centering
\includegraphics[width=0.8\columnwidth]{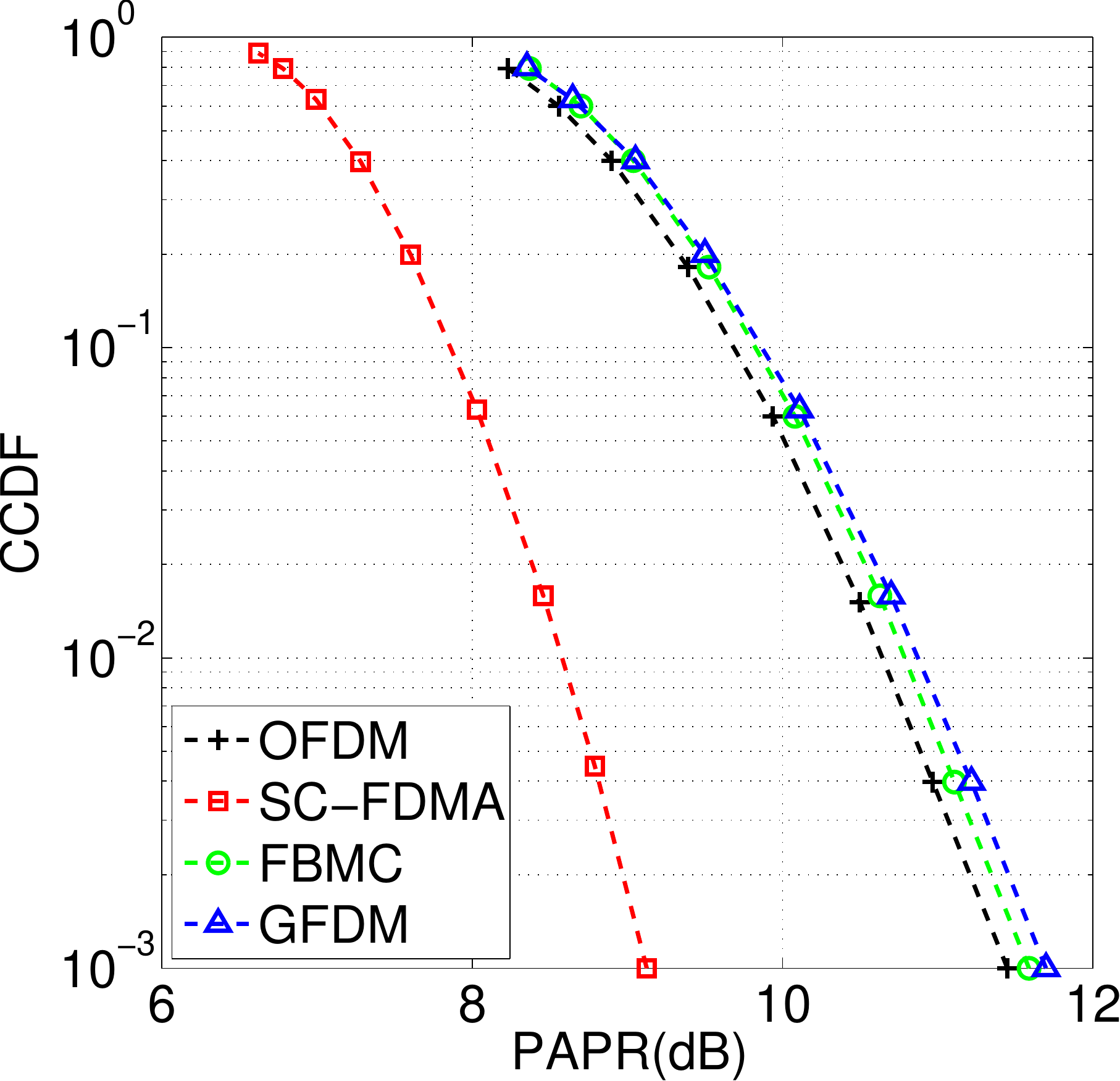}\caption{PAPR Comparison between the considered waveforms. SC-FDMA exhibits the lowest linearity requirements, whereas OFDM only slightly outperforms GFDM and FBMC.}\label{Fig:8}
\vspace{-0.3cm}
\end{figure}

We present a PAPR comparison between OFDM, SC-FDMA, GFDM and FBMC, at the transmit side in~\figref{Fig:8}. As expected, SC-FDMA has the lowest PAPR due to its inherent single-carrier structure. OFDM, GFDM, and FBMC, all being multi-carrier waveforms (with the same number of subcarriers in our setup), have similar PAPR; this  is in agreement with other studies \cite{SendreiFettweiss2014,Kollar14}. We note that in \cite{GFDMwrongpapr}, GFDM has been reported to exhibit lower PAPR than SC-FDMA since the number of subcarriers for both waveforms was not chosen to be the equal.  

\subsection{OOB Emission Performance}

\begin{figure}[tp]
\centering
\includegraphics[width=0.8\columnwidth]{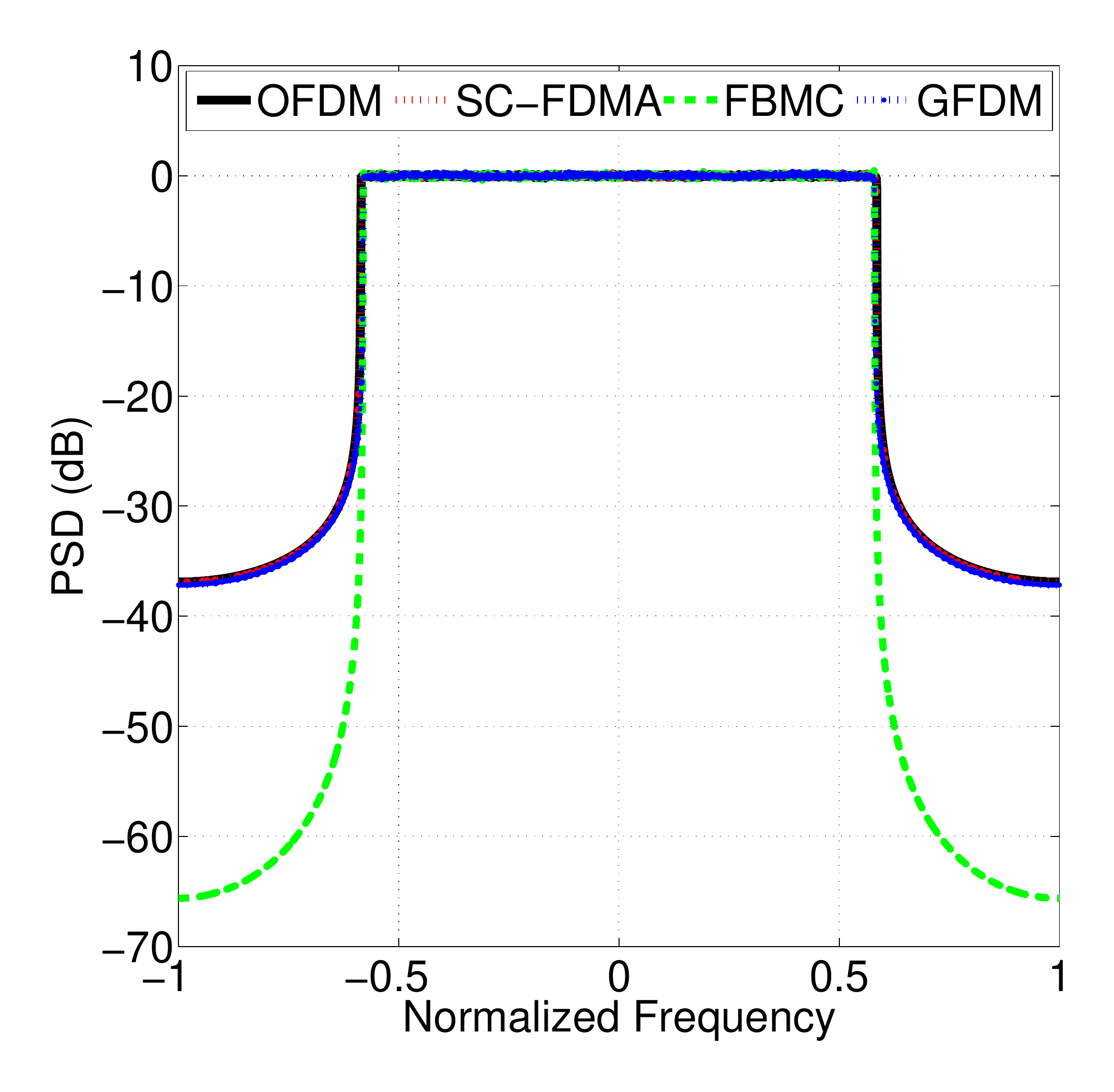}\caption{Out-of-band (OOB) leakage comparison in terms of the power spectral density (PSD). Only FBMC achieves superior OOB performance; the OOB performance plots of OFDM, SC-FDMA, and GFDM  overlap.}\label{Fig:OOB}
\vspace{-0.3cm}
\end{figure}

We present the OOB emission performance of OFDM, SC-FDMA, GFDM and FBMC, at the transmit side in~\figref{Fig:OOB}. For our setup, OFDM, SCFDMA, and GFDM exhibit identical OOB emission performance. FBMC, however, significantly outperforms all other waveforms. We believe that FBMC achieves this performance gain due to the used PHYDSAS filter which is optimized for frequency localization, whereas GFDM employs a root-raised cosine filter which is inferior to the PHYDSAS filter in terms of frequency localization. We emphasize that our results agree with those from recent studies \cite{Ucuncu15,Schellmann12} in which the same number of subcarriers was used. We also note that in \cite{GasparMMMZF15}, GFDM has been reported to exhibit better OOB emission than OFDM, which is caused by the fact that the number of subcarriers for both waveforms was not set to be equal.

\subsection{Frame Error-Rate (FER) Performance}

\begin{figure*}[tp]
\begin{center}
\includegraphics[width=1.95\columnwidth]{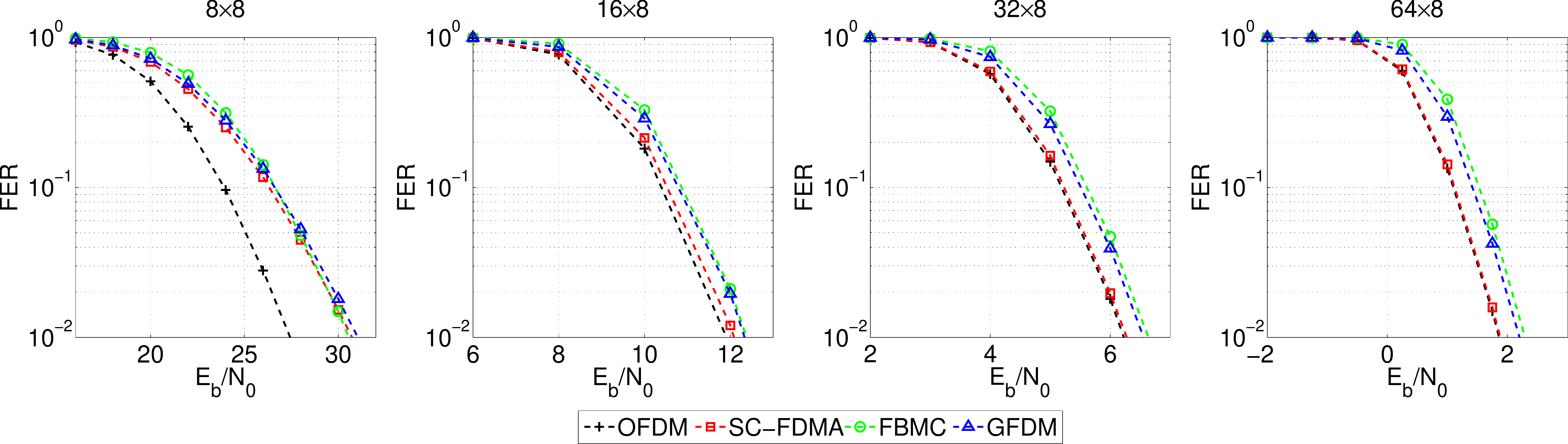}\caption{Frame error-rate (FER) comparison between the considered waveforms. OFDM achieves the best FER performance; the gap to SC-FDMA, FBMC, and GFDM decreases for large BS-to-user antenna ratios.}\label{Fig:FER}
\end{center}
\end{figure*}

In~\figref{Fig:FER}, we present a frame error-rate (FER) performance comparison between OFDM, SC-FDMA, GFDM and FBMC.
There is a large performance gap between OFDM and the other considered waveforms for a symmetric system with $8$ BS antennas and $8$ users system. For SC-FDMA, GFDM, and FBMC, significant residual inter-symbol interference (ISI) remains after demodulation, which results in a considerable large performance gap. By increasing the number of BS  antennas, the Gram matrix $\bG_{k,w}$ will become diagonally dominant due to channel hardening~\cite{larsson2014massive}; this, in turn, reduces ISI. Consequently, the performance gap of these modulation schemes becomes smaller for larger BS-to-user antenna ratios. Nevertheless, OFDM and SC-FDMA still outperform GFDM and FBMC in terms of FER. We emphasize that the presented GFDM and FBMC demodulators are suboptimal, which results in worse FER performance. The design of near-optimal but low-complexity data detectors for GFDM and FBMC remains an open research problem. 

\section{Conclusions}

In this paper, we have analyzed and compared GFDM and FBMC as potential 5G waveform candidates in the context of large-scale MIMO systems. We have proposed new, low complexity linear equalizers for both of these waveforms and compared their error-rate performance, PAPR performance, and operation count to that of OFDM and SC-FDMA. Our simulation results in a scenario with LTE-based specifications suggest that GFDM and FBMC equalized with our proposed method achieves comparable error-rate performance to MMSE-based data detection for OFDM and SC-FDMA. We have also observed that FBMC and GFDM require a considerably higher computational complexity than OFDM and SC-FDMA in large-scale MIMO systems. We have observed for our setup in which each waveform has the same number of subcarriers, the PAPR performance of GFDM and FBMC is on par with that of OFDM (but inferior to SC-FDMA). We have also observed that GFDM, OFDM, and SC-FDMA have similar OOB emission performance for our setup, whereas FBMC significantly outperforms all other waveforms. We conclude by noting that the PAPR and OOB emission performance is highly dependent on the selected parameters and studies with similar parameter settings for single-antenna systems support our findings~\cite{Kollar14,Ucuncu15,SendreiFettweiss2014}.

\bibliographystyle{IEEEtran}
\small
\bibliography{massive_MIMO_bibfile}

\end{document}